\documentclass[9pt,twocolumn,twoside]{opticajnl}
\journal{opticajournal} 

\setboolean{shortarticle}{true}

\title{Quasi-coherent perfect absorption of counter-propagating vector beams of finite spatial extent through an absorptive slab}

\author[1,*]{Sauvik Roy}
\author[1]{Nirmalya Ghosh}
\author[1]{Ayan Banerjee}
\author[1,2,3]{Subhasish Dutta Gupta}

\affil[1]{Department of Physical Sciences, IISER-Kolkata, Mohanpur 741246, India}
\affil[2]{Tata Institute of Fundamental Research, Hyderabad, Telangana 500046, India}
\affil[3]{Department of Physics, Indian Institute of Technology, Jodhpur 342030, India}

\affil[*]{sr19rs022@iiserkol.ac.in}

\begin{abstract}
Coherent perfect absorption (CPA) has been a topic of considerable contemporary research interest. 
Most of the theoretical treatment of CPA with beams, to the best of our knowledge, relies on a scalar (in some cases coupled mode) theories with inadequate input about the polarization states of the incoming light. In view of the lack of a full vectorial theory even for the original CPA configuration by Wan et al \cite{timereversedlasing_science}, we revisit the same when the incident plane waves are replaced by well defined vector beams with or without OAM.  We study the absorption characteristics of two counter-propagating monochromatic structured beams, e.g., Gaussian and Laguerre-Gaussian (LG) beams with and without orbital angular momentum, respectively, incident normally on a composite slab from both sides by fulfilling the CPA condition exclusively for the central plane wave component. We show that though perfect absorption is not achievable, there can be a substantial reduction of the scattered light. We also consider the limitations of CPA for oblique incidence and discuss the difficulties. We believe that our study will motivate and necessitate the study of recent advancements with input vector beams, retaining the full polarization information of the off-axis spatial harmonics.
\end{abstract}

\setboolean{displaycopyright}{false} 

\begin{document}

\maketitle

Coherent perfect absorption (CPA) \cite{cpa_prl, dutta_gupta_2012, dutta_gupta_2013, cpa_review_baranov,longhipra,timereversedlasing_science,gupta2015wave} is a mechanism for achieving null scattering from a scatterer under multiple illuminations. CPA has been extensively exploited across various structures, such as epsilon-near-zero (ENZ) and Kerr nonlinear materials \cite{nonlinearitysdg,NireekshanReddy,NireekshanReddyol}, metasurfaces \cite{cpa_metamaterial_selective, cpa_metamaterial_polarization}, and gratings \cite{cpa_grating_ol}, as well as in the realm of quantum optics \cite{q1referee,q2referee,agarwalcpa} with lossy beam splitters. Note that earlier works \cite{q1referee,q2referee} report precursor to CPA or CPA without naming it. CPA is the generalization of critical coupling \cite{critical_coupling_sdg,nonlinearitysdg} which involves only one input channel. An early attempt to generalize the notion of CPA to scalar beams suffer from a description of the input beams with Gaussian profile only in the plane of incidence \cite{DEY2015515} predicting inaccurate results for oblique incidence. In the context of single or multichannel inputs, there have been significant developments extending the notion of CPA to the case of any arbitrary superposition of plane waves \cite{arbitrary,ep,esurface,cpaep,arbwave}. These studies exploit exceptional points (EP) and exceptional surfaces (ES) or massively degenerate cavities requiring careful tuning of the specially prepared non-Hermitian coupled cavity systems without or with intracavity lenses. A scalar coupled mode theory was applied without adequate treatment of the polarization states of the incoming light. It is clear that the new degree of freedom introduced by the polarization of the incoming light (especially for beams carrying orbital angular momentum (OAM)) can render these advancements even richer by bringing in spin-orbit coupling effects like spin-Hall shifts into single or multichannel CPA involving both temporal and spatial frequency degeneracies. In view of the lack of a full vectorial theory, we reexamine the original CPA configuration by Wan et al \cite{timereversedlasing_science} when the incident plane waves are replaced by well-defined vector beams with or without OAM. 

\begin{figure}[h!]
\centering\includegraphics[width=1 \linewidth]{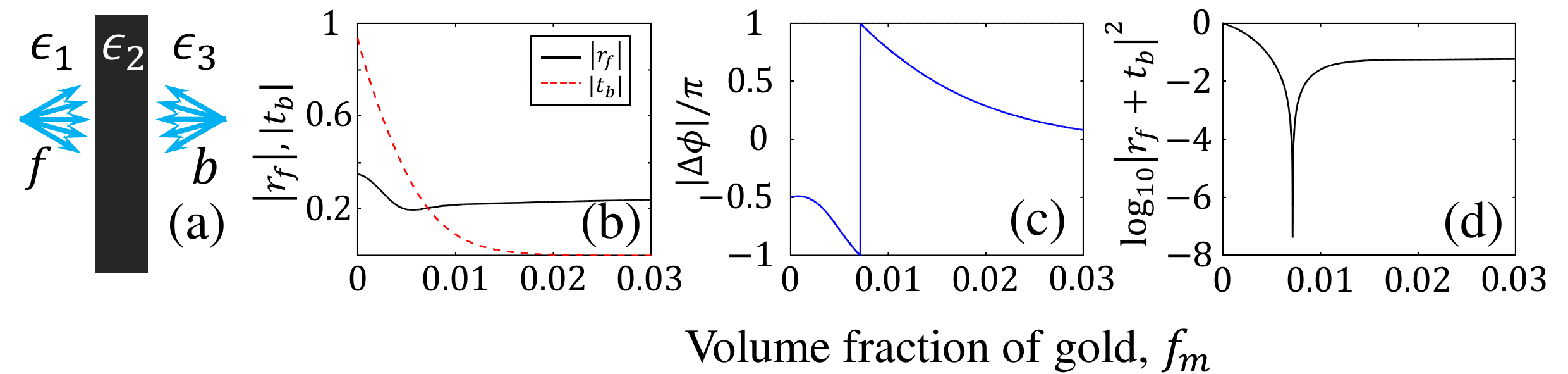}
\caption{(a) Schematic of absorption of beams with numerous plane waves incident on an absorbing slab. (b) Absolute values of the amplitude reflection and transmission coefficients $|r_f|$ and $|t_b|$ respectively. (c) The phase difference $\Delta \phi$ between the forward reflected and backward transmitted plane waves. (d) Value of $log|r_f+t_b|^2 $ as a function of volume fraction $f_m$ for the film thickness $d=4.375\mu m$ and wavelength $\lambda = 562nm$. Here, $f:$ forward propagation and $b:$ backward propagation. }
\label{cpa_plane_wave}
\end{figure}

We invoke a generic angular spectrum formalism originally developed by Bliokh and Aiello \cite{Bliokh_2013} to understand the absorption characteristics of Gaussian and  Laguerre-Gaussian (LG) beams incident symmetrically at arbitrary angles on a planar absorptive slab. We have enhanced this approach with two major improvements \cite{sounak_2024}. The strict paraxial approximation has been relaxed to accommodate illuminations with greater transverse momentum spread, and the exact Fresnel transmission $(t)$ and reflection $(r)$ coefficients have been used in place of the first-order Taylor expansion of the coefficients. All through we assumed excitation by monochromatic beams with temporal factor $e^{-i\omega t}$. Details of the formalism to calculate the reflected and transmitted beams are given in the online Supplementary Information. For convenience, we denote the forward (backward) propagating beam incident from the left (right) with subscripts '$f$' and '$b$', respectively. Since different wavevectors make different angles with the slab, it is straightforward to understand that all plane waves can not acquire the required phase difference and exhibit CPA simultaneously. This implies that there cannot be perfect absorption of the whole beam and the scattered beam will definitely have a structure. We have chosen a scenario where for the central $k$-vector, the amplitude of the reflected coefficient from the left ($|r_f|$) and the amplitude of the transmission coefficient from the right ($|t_b|$) are equal. They also possess a phase difference of $\pi$ for the $s$-polarized light and are in phase for the $p$-polarized beam. With this geometry (Fig. \ref{cpa_plane_wave}(a)), we examine various combinations of beams, incident normally on the composite slab. The most straightforward combination consists of two $s$-polarized beams: one arriving from the left and the other from the right, which is designated as the S-S combination. In contrast, the P-P combination involves two $p$-polarized beams. For circularly polarized beams, those of the same helicity can be categorized as either LCP-LCP or RCP-RCP, while those of opposite helicities can be classified as LCP-RCP or RCP-LCP. Notably, an RCP(LCP) beam under reflection becomes LCP(RCP), and thus it can interfere only with the LCP(RCP) beam incident from the other side. Therefore, for circularly polarized light, identical helicities from both sides cannot lead to destructive interference. This is clearly demonstrated in our results.  
It is worth mentioning that, for the symmetric oblique incidence, the situation becomes complex primarily due to two factors. First, it is challenging to spatially overlap the reflected beam from the left and the transmitted beam from the right due to the spatial localization of the beams. Moreover, the Goos-Hänchen (GH) and Imbert-Fedorov (IF) shifts further complicate the issue. Second, the reflection and transmission coefficients in general do not comply with the CPA condition: $r_f=t_b$ practically minimizing the possibility of destructive interference. 
\begin{figure}[h!]
\centering\includegraphics[width=1 \linewidth]{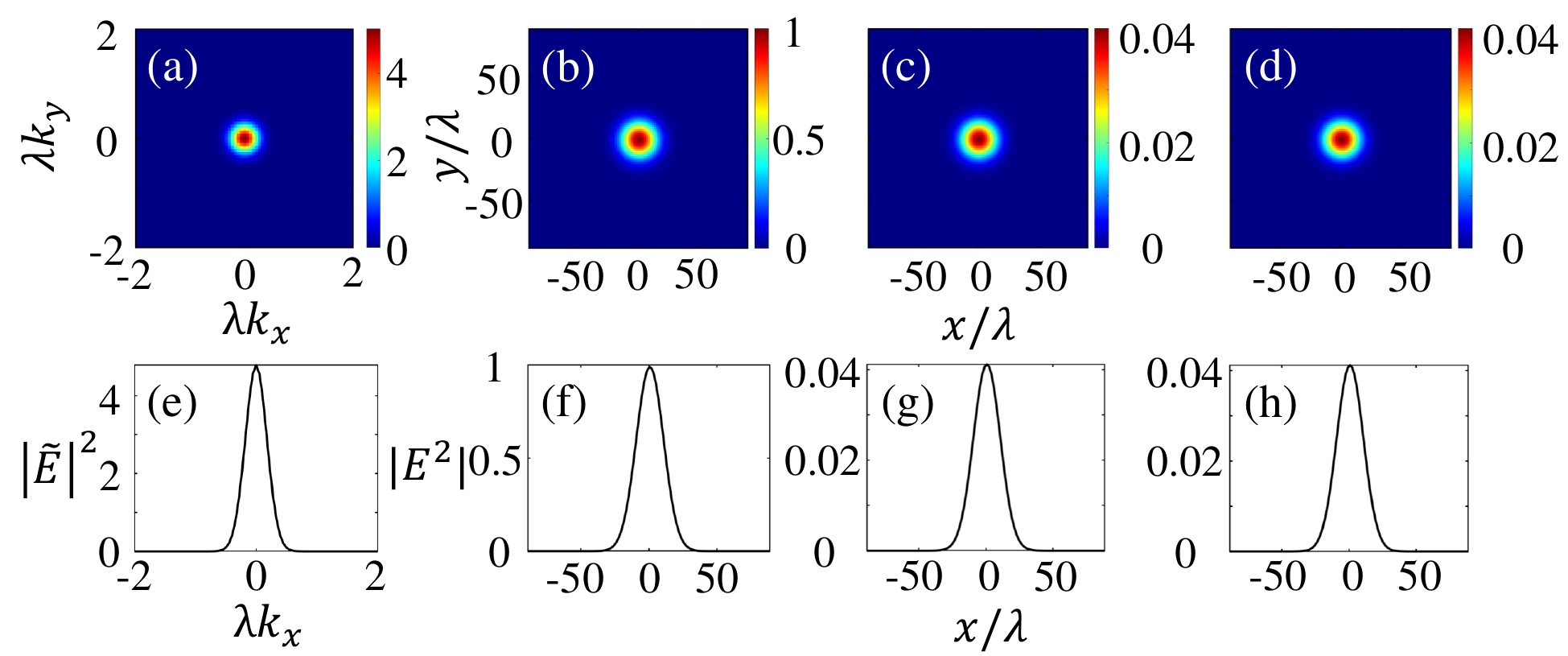}
\caption{The incident spectrum in (a) and the real-space (b) incident, (c) reflected, and (d) transmitted beams for single channel illumination from left -- all demonstrate a Gaussian character. The line plots in (e)-(h) in the second row further illustrate the Gaussian nature of the profiles in (a)-(d).}
\label{single_channel}
\end{figure}

\begin{figure}[h!]
\centering\includegraphics[width=0.9 \linewidth]{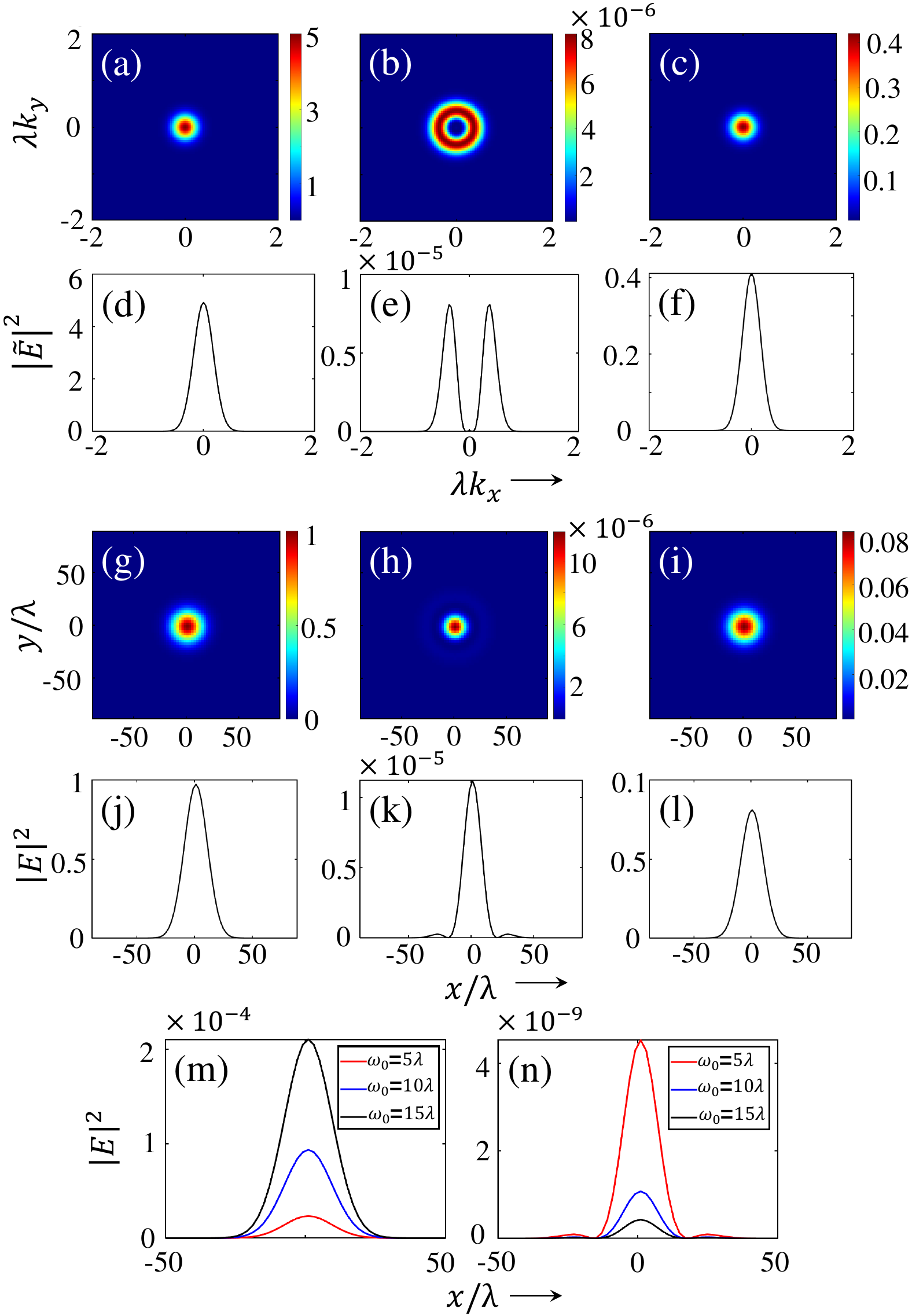}
\caption{(a)-(f) display the spectrums of the incident and the scattered beams for dual channel illumination from both sides. (a) shows the Gaussian spectrum of the incident beam, while (b) and (c) depict the scattered spectrums for S-S and LCP-LCP combinations. The line plots in (d)-(f) further illustrate the characteristics of these spectrums presented in (a)-(c). Real-space beams are presented in (g)-(l): (g) is the incident Gaussian beam, (h) shows the scattered beam for the S-S combination, and (i) represents the LCP-LCP scattered beam. Line plots (j)-(l) along the $x$-axis depict the nature of the scattered beams corresponding to (g)-(i). Finally, (m) illustrates the incident beam with three different beam waists, while (n) shows the corresponding intensities of the scattered beams. }
\label{gaussian_beam}
\end{figure}
\begin{figure}[h!]
\centering\includegraphics[width=1 \linewidth]{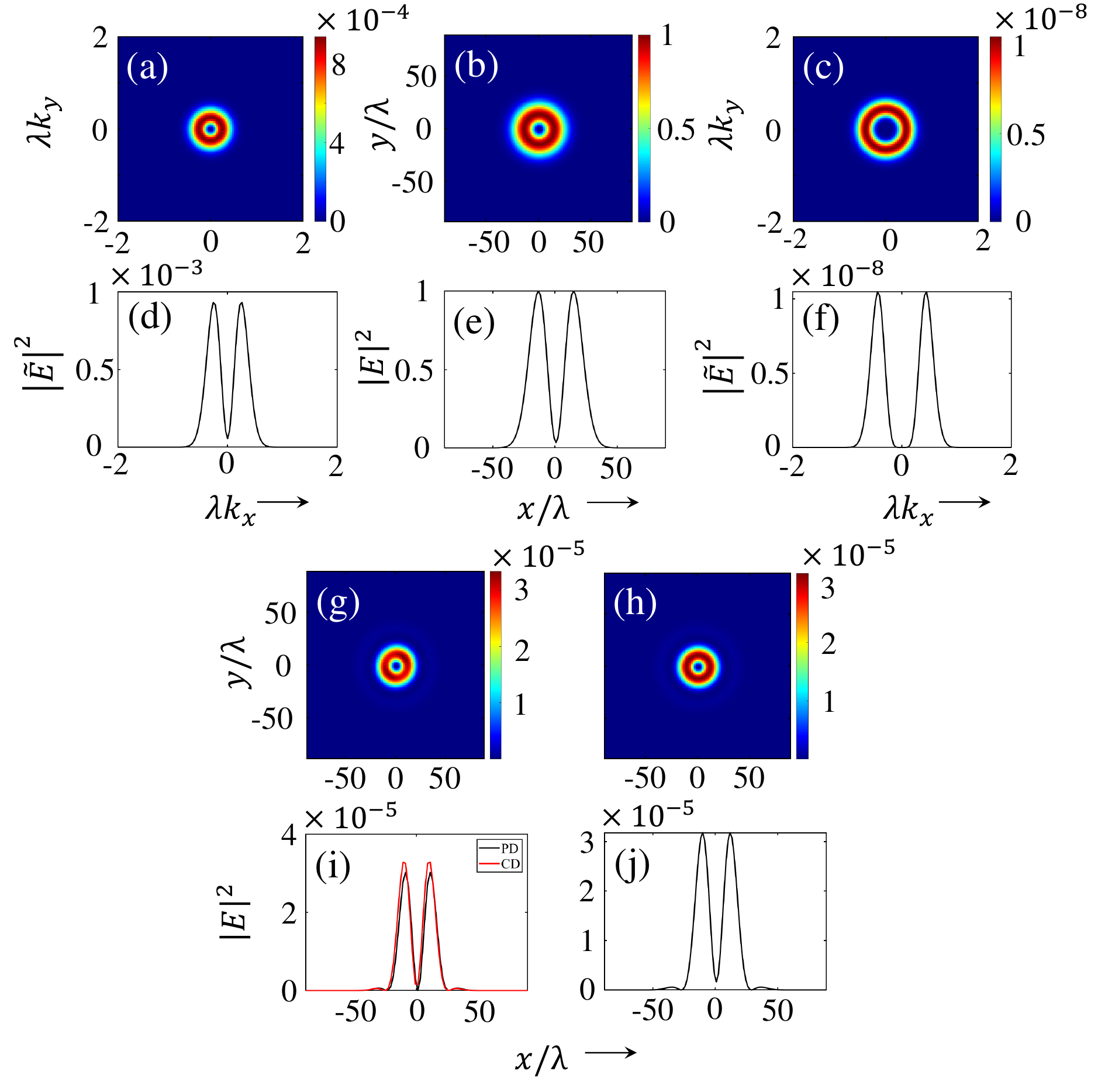}
\caption{(a) The spectrum and (b) the corresponding real LG beam with $l=1$. (c) Scattered beam spectrum for the S-S combination. The line plots in (d), (e), and (f) in the second row illustrate the characteristics of the spectrums or the beam presented in (a), (b), and (c), respectively. (g) The scattered beam for the S-S combination. The line plots in (i) along the principal diagonal (PD) and the counter diagonal (CD) highlight the nonuniformity in this scattered beam. (h) The scattered beam for the LCP-RCP combination develops a faint side lobe, which is further illustrated in (j).}
\label{lg_beam}
\end{figure}
We have chosen a gold-silica composite layer as the absorber where localized plasmon resonances determine the nature of the loss i.e., $\text{Im}(\epsilon_2)$. The dispersion characteristics of this medium $(\epsilon_2)$ are described by the Bruggeman formula \cite{optical_metamaterial, dutta_gupta_2012}: 
\begin{equation}
	\epsilon_2=\frac{1}{4} \{ p\epsilon_m + q\epsilon_d \pm \sqrt{[p \epsilon_m+q\epsilon_d]^2 + 8\epsilon_m\epsilon_d}\}
\end{equation} 
where, $p=3f_m-1$, $q=3f_d-1$, $f_m$ and $\epsilon_m$ ($f_d$ and $\epsilon_d$) are the volume fraction and permittivity of the metal (dielectric), respectively. To ensure causality, it is conventional to select the square root value that ensures a positive imaginary part of the permittivity.
\begin{figure}[h!]
\centering\includegraphics[width=0.95 \linewidth]{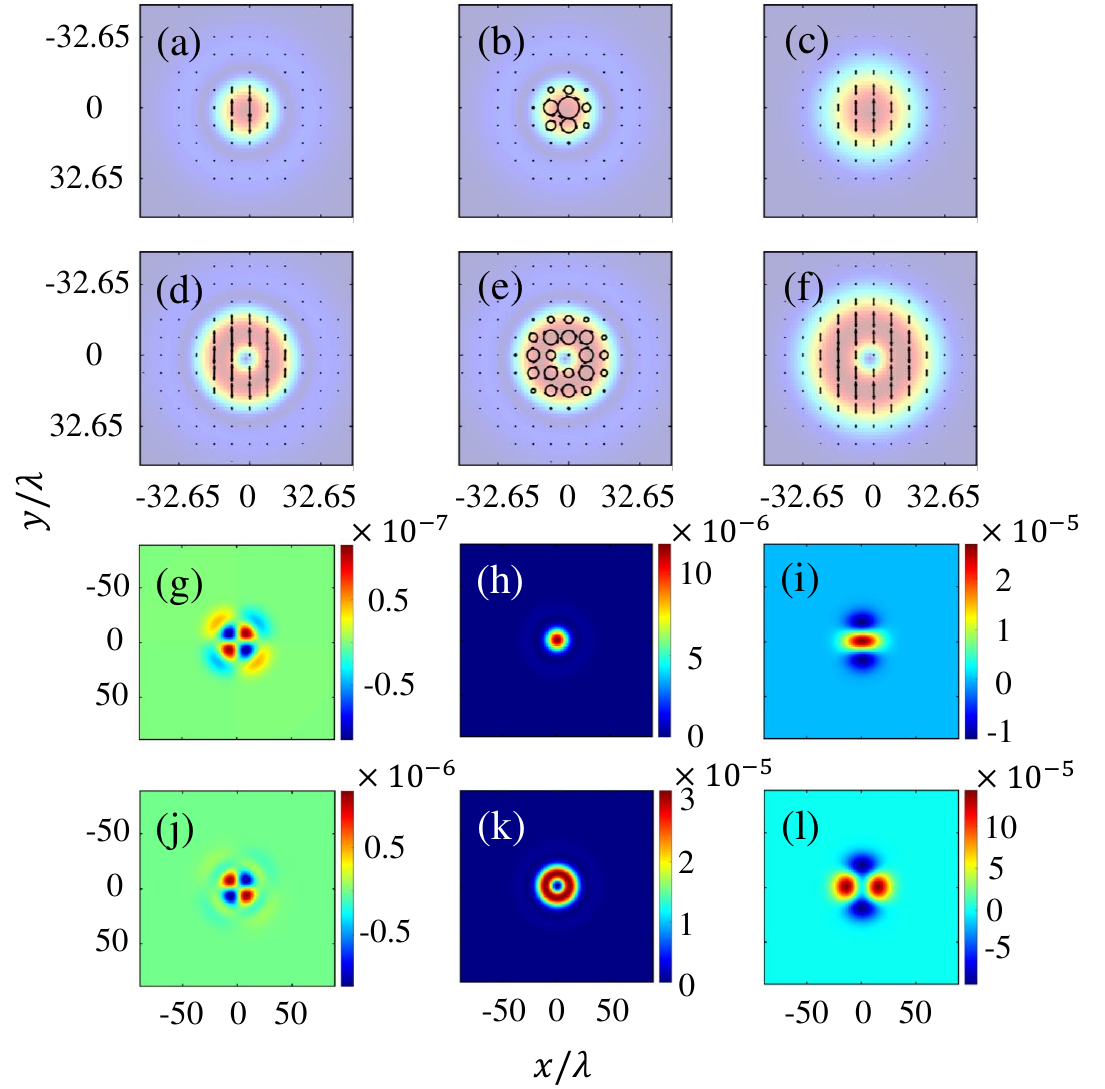}
\caption{Polarization states across the cross-section of the scattered beam for the incident (a)–(c) Gaussian beams and (d)–(f) LG beams with  $l = 1$. The beam combinations are S-S in (a) and (d), LCP-LCP in (b) and (e), and LCP-RCP in (c) and (f). The backgrounds in (a)-(f) contain contrast-enhanced and semi-transparent intensity distributions corresponding to the combinations mentioned earlier. Degree of circular polarization, $S3(=2\text{Im}(E_x^*E_y))$, across the cross-section of the scattered beam for the incident (g)–(i) Gaussian beams and (j)–(l) LG beams with $l = 1$. Again, the beam combinations are S-S in (g) and (j), LCP-LCP in (h) and (k), and LCP-RCP in (i) and (l).}
\label{polarization_state}
\end{figure}
We first look for the parameters under which the central plane wave exhibits CPA at normal incidence. The dielectric function of gold, $\epsilon_m$, was derived through an interpolation of the experimental data provided by Johnson and Christy (more in supplementary note). Other material parameters are taken as follows: $\epsilon_1=\epsilon_3=1.0$ (permittivity of free space) and $\epsilon_d = 2.25$. We plot the absolute values (Fig. \ref{cpa_plane_wave}) of the reflection, $|r_f|$, and transmission coefficients, $|t_b|$, and the phase difference between the forward and backward propagating waves as a function of the volume fraction of metal at the design wavelength of $562 \text{nm}$. It is revealed that for the thickness $d = 4.375 \mu m$, the volume fraction $f_m = 0.007$ exhibits a coincidence of $|r_f| = |t_b|$  and $|\Delta \phi| = \pi$, (Figs. \ref{cpa_plane_wave} (b), (c)) a characteristic signature of CPA for s-polarized plane waves. At this value of $f_m$, the log value of $(|r_f| + |t_b|)^2$ is found to be $-7.649$ (Fig. \ref{cpa_plane_wave} (d)) signifying a near total field suppression. It is worth noting that for a single frequency, CPA can generally be achieved across a large set of parameter values. However, the above-mentioned parameters along with beam width $w_0=10\lambda$ are used for getting the results under normal incidence. We now examine the absorption and transmission characteristics of a single beam incident on the slab, commonly referred to as single-channel illumination, and then proceed to explore the effects of dual-port or two-channel illumination. Plugging the parameters into the angular spectrum method mentioned above for a single s-polarized Gaussian beam (Figs. \ref{single_channel}(a), (b), (e), (f)) propagating from left yields that the reflected and the transmitted beams retain the Gaussian shape but with equally reduced amplitude $(|r|^2=|t|^2\approx0.04)$ (Figs. \ref{single_channel}(c), (d), (g), (h)). This can be confirmed by the near identical similarity of the patterns of the reflected (Figs. \ref{single_channel}(c), (g)) and transmitted beams (Figs. \ref{single_channel}(d), (h)) with that of the incident beam (Figs. \ref{single_channel}(b), (f)) necessary for CPA in dual illumination configuration. In short, only partial absorption of the beam takes place when a single beam interacts with the slab. 

However, in the two-port or dual-channel illumination of two s-polarized beams (i.e., S-S combination) -- although the individual beam spectra (Figs. \ref{gaussian_beam} (a), (d)) are Gaussian -- the scattered beam spectrum exhibits a dip at the center of the spectrum as depicted in Figs. \ref{gaussian_beam} (b), (e). A close examination of the line plots for the incident spectrum (Figs. \ref{gaussian_beam}(a), (d)) and the scattered spectrum with a central dip (Figs. \ref{gaussian_beam}(b), (e)) reveals that the scattered spectrum is slightly broader than the incident one. As a result, the scattered beam is observed to deviate from the Gaussian nature (Figs. \ref{gaussian_beam}(g), (j)) and develops a faint ring as seen in Figs. \ref{gaussian_beam}(h), (k). Quite obviously, the scattered beam has very little power as compared to the incident beams because of the destructive interference of the central $k$-vector(s). It is important to note that the CPA-assisted dip at the center of the spectrum is not only applicable to the S-S combinations but also to the P-P, LCP-RCP, and RCP-LCP types of illuminations. Also, the scattered beams for all these combinations are found to exhibit the same structure and power  (Figs. \ref{gaussian_beam}(h), (k)). The ratio of the maximum intensity of the incident to the scattered beam is found to be: $I^{max}_{incident}: I^{max}_{scattered} \approx 1:10^{-5}$. The only scenario where the dip does not appear is when circularly polarized beams with the same helicity—such as LCP-LCP or RCP-RCP combinations—are used (Figs. \ref{gaussian_beam}(c), (f)). These configurations do not satisfy the necessary phase and polarization relationship between the reflected and transmitted beam components, and as a result, no dip is observed in the spectrum for central plane waves. Consequently, the scattered beams have significant power and retain the Gaussian profile, as illustrated in Figs. \ref{gaussian_beam}(i), (l). Another key aspect regarding the absorption of beams is that as the beam width increases, the absorption gets enhanced as well. This is because a broad beam has less spread in the $k$-space making it closely resembling a single plane wave with a single $k$-vector. We have shown that among the three different input beam waists, $w_0 = 10\lambda$, $15\lambda$, and $20\lambda$ (Fig. \ref{gaussian_beam}(m)), the beam with the largest waist, i.e., $20\lambda$, is absorbed the most, followed by the one with $15\lambda$, and lastly, the one with $10\lambda$ (Fig. \ref{gaussian_beam}(n)). Notably, beams with different sizes on two sides are of particular interest. One such example is provided in the supplementary note.

LG beams exhibit more intriguing absorption phenomena than Gaussian beams. In this study, we examined LG beams with a vortex charge of $l = 1$ and the same polarization combinations as used with Gaussian beams. Unlike Gaussian beams, LG beams inherently possess a central dip in both real-space and $k$-space spectra (Figs. \ref{lg_beam} (a), (b), (d), (e)). Despite the zero strength of the central wave-vector, the near-destructive interference of the surrounding wave-vectors broadens the $k$-spectrum (Figs. \ref{lg_beam} (c), (f)). This broadening causes a spread in the spectra of LG beams, which, in turn, modifies the real-space profiles of the scattered beams (Figs. \ref{lg_beam} (g), (h), (i), (l)). Overall, the level of suppression is found to be of the same order of magnitude as that observed for Gaussian beams (Figs. \ref{gaussian_beam} (h), (k), \ref{lg_beam} (g), (h)).  Linearly and circularly polarized LG beams that meet the CPA requirements for central wavevectors exhibit diminished intensity, forming a faint ring in the scattered beams, similar to the Gaussian case. However, an intriguing effect is observed with linearly polarized LG beams. In the S-S combination, the uniformity of the high-intensity ring is disrupted, revealing two distinct high-intensity regions along the principal diagonal. This non-uniformity is highlighted by the line plots in (Figs. \ref{lg_beam} (g), (i)), which highlights this contrast in intensity along the principal and counter diagonals. In the P-P combination, the position of these high-intensity regions alters, appearing along the counter diagonal. This non-uniformity is a manifestation of azimuthal differential attenuation $(\mathfrak{D})$ for the orthogonal polarizations as observed in tight focusing \cite{basudebpra} and scattering \cite{ngpaper}. For the circularly polarized beams with the same helicities, the diattenuation $\mathfrak{D}$ being zero, no such nonuniformity in the scattered beams (Figs. \ref{lg_beam} (h), (j)) is observed (see supplementary).  The circularly polarized beams with the same helicity exhibit a similar type of nature as in the case of the Gaussian beam. Neither the spectrum gets altered nor is there any modification in the real space beams. These beams get absorbed in the slab to the same extent as a single beam and get simply added up in the scattered beam because of the two-port illumination scheme. Another key factor is the polarization state across the cross-section of the scattered beam. Due to interactions with the slab, the polarization generally loses its uniformity across the beam. For instance, in the case of linearly polarized beams -- while the overall polarization remains largely linear -- a degree of circular polarization arises. This results from the partial interconversion between the $s$- and $p$-polarization components during the beam's interaction with the slab. This phenomenon is the same for both the linearly polarized Gaussian beam (Fig. \ref{polarization_state} (a)) and for the linearly polarized LG beam (Fig. \ref{polarization_state} (d)). The amount of circular polarization introduced is displayed in Fig. \ref{polarization_state} (g) for the Gaussian beam and in Fig. \ref{polarization_state} (j) for the LG beam. In this case, typically, four lobes with opposite circular polarization, again dictated by the variation in diattenuation $\mathfrak{D}$, can be observed across the beam. On the contrary, for the circularly polarized beams, a predominantly circular state of polarization is found across the entire cross-section. For the LCP-RCP (RCP-LCP) type of combination, only the positive (negative) helicity regions can be seen across the scattered beam. Interestingly, for the circularly polarized light with the same helicity, the $p$-component of the polarization gets canceled and predominantly the $s$-polarization is left in the scattered beam. In this case, both positive and negative helicity regions can also be found juxtaposed in both Gaussian and LG beams (Figs. \ref{polarization_state} (i), (l)). 

In conclusion, we have extended the work of \cite{timereversedlasing_science} on CPA to include the effect of polarized input beams. Using the angular spectrum method, we examined the impact of beam polarization when interacting with an absorptive slab under both single and dual-channel illumination. Our findings reveal that complete absorption of a beam is not achievable within this configuration, highlighting a limitation of the CPA mechanism specific to this platform. However, our approach offers potential for further exploration, such as extending the analysis to oblique incidence, illumination on photonic crystals, metasurfaces, nonlinear medium, or low-power applications. Moreover, the structured nature of the scattered beam opens up new possibilities for beam shaping. This could be achieved by employing two different beams incident from opposite sides, or by cascading the output of one absorptive slab to serve as the input for another, thus expanding the scope of CPA applications. We believe that our study will motivate and necessitate the study of recent developments on CPA with ES, EP, and degenerate cavities with full vectorial beams.

\begin{backmatter}
\bmsection{Acknowledgment} Sauvik Roy is thankful to the Department of Science and
Technology (DST), Government of India, for the INSPIRE fellowship.
\bmsection{Disclosures} The authors declare no conflicts of interest.
\bmsection{Data Availability Statement} Data underlying the results presented in this paper are
not publicly available at this time but may be obtained from the authors upon reasonable request.
\bmsection{Supplemental document}See Supplement 1 for supporting content.
\end{backmatter}

\begin{thebibliography}{10}
\newcommand{\enquote}[1]{``#1''}

\bibitem{timereversedlasing_science}
W.~Wan, Y.~Chong, L.~Ge, \emph{et~al.}, \enquote{Time-reversed lasing and interferometric control of absorption,} {\protect\JournalTitle{Science}} \textbf{331}, 889--892 (2011).

\bibitem{cpa_prl}
Y.~D. Chong, L.~Ge, H.~Cao, and A.~D. Stone, \enquote{Coherent perfect absorbers: Time-reversed lasers,} {\protect\JournalTitle{Phys. Rev. Lett.}} \textbf{105}, 053901 (2010).

\bibitem{dutta_gupta_2012}
S.~Dutta-Gupta, O.~J.~F. Martin, S.~D. Gupta, and G.~S. Agarwal, \enquote{Controllable coherent perfect absorption in a composite film,} {\protect\JournalTitle{Opt. Express}} \textbf{20}, 1330--1336 (2012).

\bibitem{dutta_gupta_2013}
K.~N. Reddy and S.~D. Gupta, \enquote{Light-controlled perfect absorption of light,} {\protect\JournalTitle{Opt. Lett.}} \textbf{38}, 5252--5255 (2013).

\bibitem{cpa_review_baranov}
D.~G. Baranov, A.~Krasnok, T.~Shegai, \emph{et~al.}, \enquote{Coherent perfect absorbers: linear control of light with light,} {\protect\JournalTitle{Nature Reviews Materials}} \textbf{2}, 1--14 (2017).

\bibitem{longhipra}
S.~Longhi, \enquote{Coherent perfect absorption in a homogeneously broadened two-level medium,} {\protect\JournalTitle{Phys. Rev. A}} \textbf{83}, 055804 (2011).

\bibitem{gupta2015wave}
S.~D. Gupta, N.~Ghosh, and A.~Banerjee, \emph{Wave optics: Basic concepts and contemporary trends} (CRC Press, 2015).

\bibitem{nonlinearitysdg}
K.~N. Reddy, A.~V. Gopal, and S.~D. Gupta, \enquote{Nonlinearity induced critical coupling,} {\protect\JournalTitle{Opt. Lett.}} \textbf{38}, 2517--2520 (2013).

\bibitem{NireekshanReddy}
K.~N. Reddy and S.~D. Gupta, \enquote{Gap solitons with null-scattering,} {\protect\JournalTitle{Opt. Lett.}} \textbf{39}, 2254--2257 (2014).

\bibitem{NireekshanReddyol}
K.~N. Reddy and S.~D. Gupta, \enquote{Light-controlled perfect absorption of light,} {\protect\JournalTitle{Opt. Lett.}} \textbf{38}, 5252--5255 (2013).

\bibitem{cpa_metamaterial_selective}
G.~Nie, Q.~Shi, Z.~Zhu, and J.~Shi, \enquote{{Selective coherent perfect absorption in metamaterials},} {\protect\JournalTitle{Applied Physics Letters}} \textbf{105}, 201909 (2014).

\bibitem{cpa_metamaterial_polarization}
M.~Kang, F.~Liu, T.-F. Li, \emph{et~al.}, \enquote{Polarization-independent coherent perfect absorption by a dipole-like metasurface,} {\protect\JournalTitle{Opt. Lett.}} \textbf{38}, 3086--3088 (2013).

\bibitem{cpa_grating_ol}
S.~Dutta-Gupta, R.~Deshmukh, A.~V. Gopal, \emph{et~al.}, \enquote{Coherent perfect absorption mediated anomalous reflection and refraction,} {\protect\JournalTitle{Opt. Lett.}} \textbf{37}, 4452--4454 (2012).

\bibitem{q1referee}
S. ~M. Barnett, J. ~Jeffers, A. ~Gatti, R. ~Loudon, \enquote{Quantum optics of lossy beam splitters,} {\protect\JournalTitle{Phys. Rev. A}} \textbf{57}, 2134 (1998).

\bibitem{q2referee}
J. ~Jeffers, \enquote{Interference and the lossless lossy beam splitter} {\protect\JournalTitle{Journal of Modern Optics}} \textbf{47}, 1819 (2000).

\bibitem{agarwalcpa}
S.~Huang and G.~S. Agarwal, \enquote{Coherent perfect absorption of path entangled single photons,} {\protect\JournalTitle{Opt. Express}} \textbf{22}, 20936--20947 (2014).

\bibitem{critical_coupling_sdg}
S.~D. Gupta, \enquote{Strong-interaction---mediated critical coupling at two distinct frequencies,} {\protect\JournalTitle{Opt. Lett.}} \textbf{32}, 1483--1485 (2007).

\bibitem{DEY2015515}
S.~Dey, \enquote{Coherent Perfect Absorption using Gaussian beams,} {\protect\JournalTitle{Optics Communications}} \textbf{356}, 515--521 (2015).

\bibitem{arbitrary}
Y.~Slobodkin, G.~Weinberg, H.~Hörner,  \emph{et~al.}, \enquote{Massively degenerate coherent perfect absorber for arbitrary wavefronts,} {\protect\JournalTitle{Science}} \textbf{377}, 995 (2022).

\bibitem{ep}
W.~R. Sweeney, C.~W Hsu, S.~Rotter, \emph{et~al.}, \enquote{Perfectly Absorbing Exceptional Points and Chiral Absorbers,} {\protect\JournalTitle{Phys. Rev. Lett.}} \textbf{122}, 093901 (2019).

\bibitem{esurface}
S.~Soleymani, Q.~Zhong, M.~Mokim, \emph{et~al.}, \enquote{Chiral and degenerate perfect absorption on exceptional surfaces,} {\protect\JournalTitle{Nature Communications}} \textbf{13}, 599 (2022).

\bibitem{cpaep}
C.~Wang, W.~R. Sweeney, A.~Douglas Stone,  \emph{et~al.}, \enquote{Coherent perfect absorption at an exceptional point,} {\protect\JournalTitle{Science}} \textbf{373}, 1261--1265 (2021).

\bibitem{arbwave}
H.~H\"orner, L.~Wild,  and W.~Yevgeny, \emph{et~al.}, \enquote{Coherent Perfect Absorption of Arbitrary Wavefronts at an Exceptional Point,} {\protect\JournalTitle{Phys. Rev. Lett.}} \textbf{133}, 173801 (2024).

\bibitem{Bliokh_2013}
K.~Y. Bliokh and A.~Aiello, \enquote{Goos–hänchen and imbert–fedorov beam shifts: an overview,} {\protect\JournalTitle{Journal of Optics}} \textbf{15}, 014001 (2013).

\bibitem{sounak_2024}
S.~S. Biswas, G.~Remesh, V.~G. Achanta, \emph{et~al.}, \enquote{Reflectionless propagation of beams through a stratified medium,} {\protect\JournalTitle{Optics Communications}} \textbf{569}, 130766 (2024).

\bibitem{optical_metamaterial}
W.~Cai and V.~Shalaev, \enquote{Optical metamaterials: Fundamentals and applications,} {\protect\JournalTitle{Springer New York, NY}}  (2010).

\bibitem{basudebpra}
B.~Roy, N.~Ghosh, S.~Dutta~Gupta, \emph{et~al.}, \enquote{Controlled transportation of mesoscopic particles by enhanced spin-orbit interaction of light in an optical trap,} {\protect\JournalTitle{Phys. Rev. A}} \textbf{87}, 043823 (2013).

\bibitem{ngpaper}
N.~Ghosh, A.~I. Vitkin, \enquote{Tissue polarimetry: concepts, challenges, applications, and outlook,} {\protect\JournalTitle{Journal of Biomedical Optics}} \textbf{16}, 110801 (2011).



\end{thebibliography}

\providecommand{\noopsort}[1]{}\providecommand{\singleletter}[1]{#1}%



\providecommand{\noopsort}[1]{}\providecommand{\singleletter}[1]{#1}%


\end{document}